# Realization of Epitaxial Thin Films of the Topological Crystalline Insulator Sr$_3$SnO

Yanjun Ma[1,2], Anthony Edgeton[2], Hanjong Paik[3], Brendan Faeth[4], Chris Parzyck[4], Betül Pamuk[3], Shun-Li Shang[5], Zi-Kui Liu[5], Kyle M. Shen[4,6], Darrell G. Schlom[6,7] and Chang-Beom Eom[2*]

[1]*Department of Physics and Astronomy, West Virginia University, Morgantown, West Virginia 26506, USA*

[2]*Department of Material Science and Engineering, University of Wisconsin-Madison, Madison, Wisconsin 53706, USA*

[3]*Platform for the Accelerated Realization, Analysis, and Discovery of Interface Materials (PARADIM), Cornell University, Ithaca, New York 14853, USA*

[4]*Laboratory of Atomic and Solid State Physics, Cornell University, Ithaca, New York 14853, USA*

[5]*Department of Materials Science and Engineering, The Pennsylvania State University, University Park, Pennsylvania 16802, USA*

[6]*Kavli Institute at Cornell for Nanoscale Science, Ithaca, New York 14853, USA*

[7]*Department of Material Science and Engineering, Cornell University, Ithaca, New York 14853, USA*

**Topological materials are derived from the interplay between symmetry and topology. Advances in topological band theories have led to the prediction that the antiperovskite oxide Sr$_3$SnO is a topological crystalline insulator, a new electronic phase of matter where the conductivity in its (001) crystallographic planes is protected by crystallographic point group symmetries. Realization of this material, however, is challenging. Guided by thermodynamic calculations we design and implement a deposition approach to achieve the adsorption-controlled growth of epitaxial Sr$_3$SnO single-crystal films by molecular-beam epitaxy (MBE). *In-situ* transport and angle-resolved photoemission spectroscopy measurements reveal the metallic and non-trivial topological nature of the as-grown samples. Compared with conventional MBE, the synthesis route used results in superior sample quality and is readily adapted to other topological systems with antiperovskite structures. The successful realization of thin films of topological crystalline insulators opens opportunities to manipulate topological states by tuning symmetries via epitaxial strain and heterostructuring.**

---

* Correspondence should be sent to eom@engr.wisc.edu.

**Introduction**

The fundamental principle for understanding matter has its roots in symmetry and topology[1]. An outstanding example is the development of topological insulators (TIs), where the interaction between time-reversal symmetry and topology of the electron wave function drives the emergence of a dissipationless metallic boundary state[2,3]. Inspired by the discovery of topological insulators, the classification of topological phases has been extended by including other symmetry classes such as particle-hole symmetry that results in topological superconductors[4-6] and crystallographic point group symmetries that lead to topological crystalline insulators (TCIs)[7]. TCIs exhibit a variety of novel topological electronic properties such as Dirac mass generation via ferroelectric distortion[8,9] and strain-induced flat band superconductivity[10]. Given these developments, there is great interest in identifying TCI materials and proposals for them include pyrochlore iridates[11], heavy-fermion compounds[12] and antiperovskite oxides $A_3BO$ (space group $Pm\bar{3}m$ with $A$ = Sr, Ca and Ba, $B$ = Pb and Sn, O is oxygen)[13]. TCIs have been first observed in various bulk crystals including SeTe[14], $Pb_{1-x}Sn_xSe$[15], $Pb_{1-x}Sn_xTe$[16] and $Ca_3PbO$[17].

Compared with TIs, the surface states of TCIs have a much wider range of tunable electronic properties under various perturbations including structural distortion, mechanical strain and thickness control, which makes thin-film TCIs promising for topological electronics[18,19]. So far, the investigation on thin-film TCIs have been focused on SnSe- or SnTe-based compounds[20-25]. While they exhibit non-trivial topological properties, the chemical nature of chalcogenides makes the integration of such materials with multi-functional oxides, and hence the delivery of applicable devices, challenging[26,27]. In addition, the lattice constants for SnSe- and SeTe-based materials are larger than 6 Å, which renders the identification of suitable substrates for optimizing film qualities difficult[20,21]. In contrast, as proposed by theories, owing to their chemistries and crystal structures, topological oxide materials are more promising for developing device concepts[11,13]. To advance the realization of topological electronics, a thin-film topological oxide material needs to be identifed[28].

The antiperovskite oxide $Sr_3SnO$ (Fig. 2a) is not only a TCI candidate, but also potentially superconducting with $T_c \approx 4$ K[29-31], making this material an ideal system to investigate topological superconductivity. The synthesis of this antiperovskite is, however, challenging because the unusual valence state of Sn makes the growth of $Sr_3SnO$ much less forgiving than its perovskite counterparts[32]. The deposition of epitaxial thin films of $Sr_3PbO$, a sibling to $Sr_3SnO$, has been realized recently by molecular-beam epitaxy (MBE)[33]. Transport measurement showed that the samples were doped with holes, complicating the observation of Dirac features. To enhance sample quality, we have employed a different growth approach.

## Results

**Synthesis of thin-film Sr$_3$SnO by design**

Adsorption-controlled growth has proven to be extremely effective for realizing accurate and reproducible control of film stoichiometry[34-36]. Recent advances in materials chemistry and thermodynamic modelling relevant to synthesis have enabled remarkably accurate guidance for engineering various complex oxides by adsorption-controlled growth[37-39]. To produce more optimal TCI thin films, we investigated thermodynamic conditions amenable to the adsorption-controlled growth of Sr$_3$SnO using a combination of thermodynamic data from the literature together with density functional theory (DFT) and phonon calculations for Sr$_3$SnO (see Fig. 1 and the Supplementary Information for more details). In our computational model, by considering the equilibrium states of strontium and the balance of the reaction between Sr and SnO, we identified the adsorption-controlled growth regime for generating phase-pure Sr$_3$SnO (the shaded area in Fig. 1). To produce the desired growth conditions in MBE experiments, a molecular beam of Sr was produced by evaporating Sr from an effusion cell while a SnO molecular beam was formed through the thermal decomposition of SnO$_2$ contained within an effusion cell at elevated temperature[39,40]. The proposed growth window was validated by experiments; the conditions of all the synthesized single-crystal Sr$_3$SnO samples have been found in the predicted range (blue markers in Fig. 1).

**Experimental realization and structural characterization**

Sr$_3$SnO thin films with [001] out-of-plane direction were deposited on (001)-oriented yttrium-stabilized ZrO$_2$ (YSZ) single-crystal substrates. Reflection high-energy electron diffraction (RHEED) was employed to monitor the growth in real time. By carefully tuning the substrate temperature, we achieved streaky RHEED patterns along Sr$_3$SnO [100] and [110] azimuths, indicating films with smooth surfaces and high crystalline quality (Fig. 2b and see the Supplementary Information for more discussion of RHEED results). Because Sr$_3$SnO is extremely reactive in air[32,33], *in-situ* Au capping was utilized to protect the as-grown samples. The Au/Sr$_3$SnO/YSZ heterostructure was analysed *ex-situ* by X-ray diffraction (XRD) following capping. Symmetric $\theta$-$2\theta$ XRD scans detected odd-order Sr$_3$SnO $00\ell$ peaks, while even-order $00\ell$ peaks were masked by those from YSZ because Sr$_3$SnO and YSZ are closely lattice matched (~0.5%) (Fig. 2c). Together, the result in Fig. 2c and the $\phi$-scan in Fig. 2d confirmed the "cube-on-cube" orientation relationship between Sr$_3$SnO thin films and the YSZ substrate, i.e., $[100]_{Sr_3SnO} \parallel [100]_{YSZ}$ and $(001)_{Sr_3SnO} \parallel (001)_{YSZ}$.

It is known from previous work that impurity phases such as SrO can be formed during synthesis[33]. To investigate the phase purity of the grown films, we utilized (001)-oriented LaAlO$_3$ (LAO) single-crystal substrates (lattice mismatch ~ 4%) so that the XRD peaks of the film could be well distinguished from those of the substrate. Epitaxial Sr$_3$SnO films were formed on LAO under similar conditions as on YSZ (see the Supplementary Information for more discussions on the growth of Sr$_3$SnO on LAO). We performed asymmetric $\theta$-$2\theta$ scans of the films grown on LAO substrates. As discussed in the literature[30], if the Sr$_3$SnO film is not phase pure and also contains

SrO, one should observe a double-peak in the $\theta$-$2\theta$ scan of the Sr$_3$SnO 113 and 202 peaks. In our high-resolution $\theta$-$2\theta$ scans, however, only a single peak was detected as is evident in Fig. 2e and f. This confirms that our Sr$_3$SnO films are free of SrO.

*In-situ* **investigation**

In order to probe the electronic structure and properties of the Sr$_3$SnO thin films to assess the predicted topological crystalline insulator nature[13,41], we performed angle-resolved photoemission spectroscopy (ARPES) and four-point electrical measurements in the van der Pauw geometry. Due to the extreme reactivity of Sr$_3$SnO with air, all measurements were performed on uncapped films in ultrahigh vacuum, without ever exposing the Sr$_3$SnO films to pressures higher than 2 x 10$^{-10}$ torr. The as-grown sample exhibited metallic behavior with a sheet resistance in the range of $10^2$ $\Omega/\square$; at temperatures below 30 K, a small upturn was observed (Fig. 3a). The metallicity of the Sr$_3$SnO films agreed with previous reports on bulk Sr$_3$SnO crystals[30], and was similar to the transport properties of bulk Ca$_3$PbO[42] and MBE-grown Sr$_3$PbO films[33]. Measurements on four other samples also exhibit sheet resistances in the range of $10^2 \sim 10^3$ $\Omega/\square$ and metallic behavior between 30 and 300 K, with some of the less metallic samples demonstrating signatures of localization below 30 K (see the Supplemental Information).

The surface crystal structure of the as-grown films was evaluated by *in-situ* low-energy electron diffraction (LEED) (see Fig. 3b and the Supplementary Information for more details). Careful analysis of the LEED results revealed the cubic symmetry with a lattice constant of $a \approx$ 5.14 Å in agreement with the reported value in the literature ($a$ = 5.1394 Å)[32]. LEED showed that the as-grown surface does not exhibit any surface reconstruction. All samples were checked by LEED to assure that the properties measured in our experiments are intrinsic to Sr$_3$SnO.

In Fig. 3c, ARPES spectra from the same sample measured in Fig. 3a are shown along the in-plane high symmetry directions (He $I\alpha$ $h\nu$ = 21.2 eV, $T$ = 15 K), together with a comparison with DFT calculations of Sr$_3$SnO (PBEsol + SOC) at an out-of-plane momentum of $k_z$ = 0.7 $\pi/c$, corresponding to the $k_z$ value measured at He $I\alpha$ $h\nu$ = 21.2 eV), which was determined by comparing the ARPES spectra across a range of $k_z$ values (see the Supplemental Information). The DFT calculations and ARPES spectra show remarkably close agreement, confirming that the electronic structure of Sr$_3$SnO agrees well with the predicted TCI electronic structure[13,41]. Furthermore, the close agreement between the ARPES spectra and DFT calculations near the $\Gamma$ point (0,0,0.35) is particularly noteworthy, given that the touching of this band at $\Gamma$ is essential for governing the TCI nature of Sr$_3$SnO. Our detailed investigation into the orbitally projected electronic band structure is consistent with the non-trivial topological nature, specifically the crossing of the Sr *4d* and Sn *5p* states near the $\Gamma$ point[13,29] (See the Supplementary Information for details). Finally, we note that the calculated electronic structure must be shifted downwards by 0.6 eV in order to match the experimental ARPES data, corresponding to an electron doping of 0.2$e^-$ / Sn. The reason for this shift of the band structure is not yet clear but might arise from the polar nature of the Sr$_3$SnO (001) structure (alternating layers of [Sr$_2$O]$^{2+}$ and [SrSn]$^{2-}$), or from inadvertent oxygen vacancies. This is being investigated further.

In summary we have utilized thermodynamic modelling to identify an adsorption-controlled growth regime for single-phase films of the TCI Sr$_3$SnO. This synthesis approach leads to films of exceptional quality enabling the intrinsic non-trivial topology of the correlated electrons in Sr$_3$SnO by ARPES to be established. The fabrication process can be adapted to produce other *A$_3$B*O-based TCI materials, which will significantly expand the materials space for TCI research and foster the engineering of topological electronics.

**Methods:**

**Preparation of YSZ and LAO substrates for thin film deposition experiments**

All LAO (001) and YSZ (001) substrates used in the experiments were commercially available from CrysTec GmbH. The back side of each substrate was coated with 200 nm Pt layer that enhanced the efficiency of radiation heating. All substrates were first cleaned with acetone and isopropyl alcohol (IPA) and then rinsed with deionized water. For LAO (001) single crystals, thermal annealing in the air at 1100 °C for 10 hours were carried out. After annealing, the substrates were boiled in water for 15 minutes to establish AlO$_2$-terminated surfaces that exhibited regular terraces with single unit cell step height.

**Molecular beam epitaxy growth of Sr$_3$SnO**

Prefilled distilled Sr (purity = 99.99%) in titanium crucibles from Sigma Aldrich (Millipore Sigma) and SnO$_2$ powder (purity = 99.996%) from Alfa Aesar (Puratronic®) were used as source materials. The base pressure of the growth chamber was below $2 \times 10^{-9}$ Torr. During the deposition, the background pressure was on the order of 10$^{-8}$ Torr. The substrate temperature was monitored by a thermocouple on the back of the sample holder and a pyrometer pointing to the front surface of the substrate. Before each deposition, the flux of source materials was measured by a quartz crystal microbalance.

***In-situ* transport and ARPES measurement**

To avoid sample degradation films are grown and characterized within a single integrated UHV manifold (p<2E-10 Torr) equipped with ARPES, LEED, and an *in-situ* transport. Pristine films were contacted in UHV for van der Pauw four-point probe transport measurements via spring loaded probes attached to a custom-built cryostat with a base temperature of 5K. DC resistance measurements were taken using a Keithley 6221/2182A current source/voltmeter combination with a typical applied current of 1-10uA. ARPES measurements were taken with a Scienta R4000 electron analyzer equipped with a VUV5000 helium discharge lamp using He-I photons at 21.2 eV.

**Density functional theory calculations and thermodynamic computations**

The details regarding the theoretical calculations are given in the Supplementary Information.

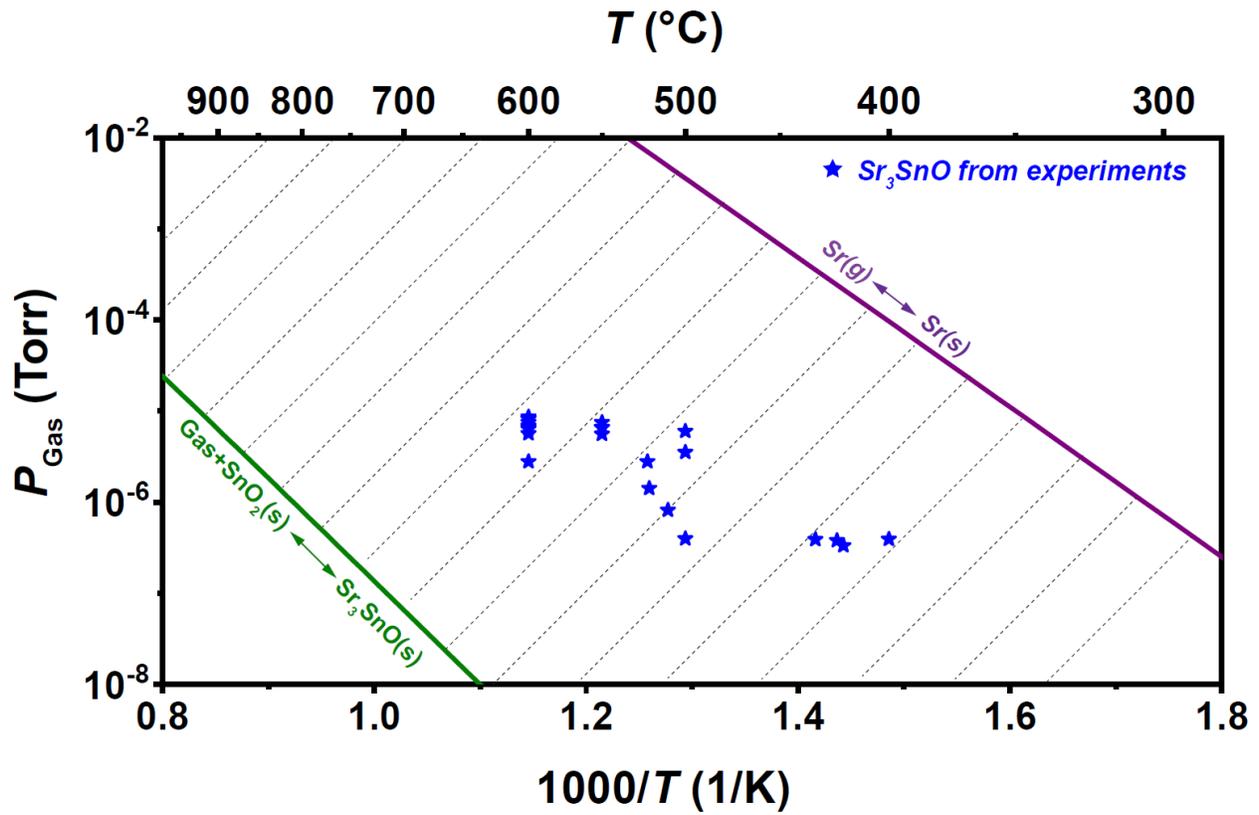

**Fig. 1** Computed phase stability diagram for the Sr-Sn-O system. The shaded area represents the adsorption-controlled growth regime for the Sr$_3$SnO phase based on calculations, while the blue markers represent the conditions at which single-crystal Sr$_3$SnO was synthesized experimentally.

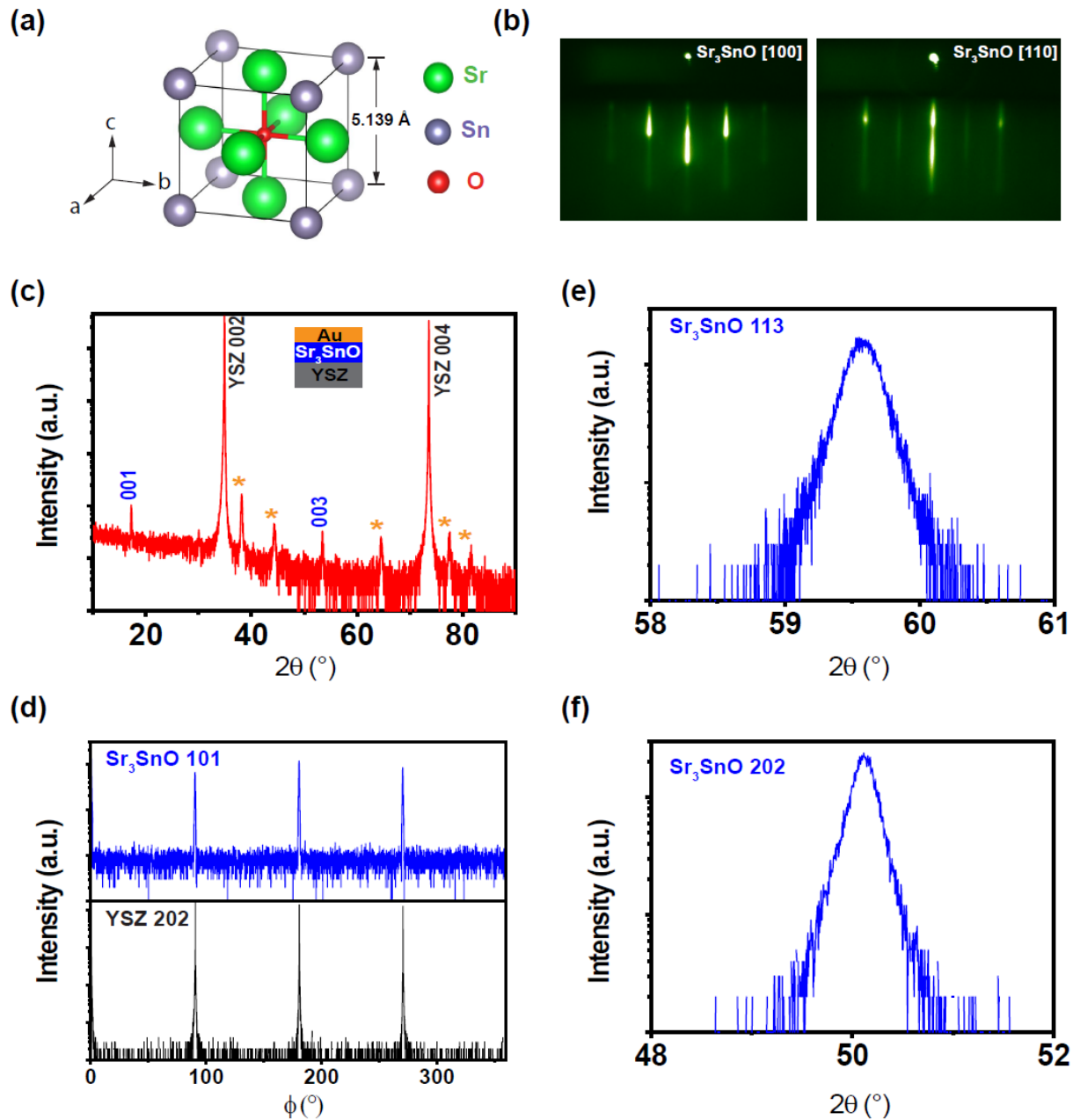

**Fig. 2** Structural characterization of $Sr_3SnO$ thin films. **a** Crystalline structure of $Sr_3SnO$. **b** RHEED patterns for the as-grown $Sr_3SnO$ films viewed along [100] and [110] azimuths, respectively. **c** Symmetric $\theta$-$2\theta$ XRD scan shows odd order $00\ell$ reflections from the $Sr_3SnO$ film. The reflections from the Au capping layer are marked with asterisks. **d** $\phi$-scan demonstrates the in-plane epitaxial relationship between the $Sr_3SnO$ film and underlying YSZ substrate. **e** and **f** Asymmetric $\theta$-$2\theta$ XRD scans from $Sr_3SnO$ 113 and 202, respectively. The single-peak feature confirms that there is no SrO present in the sample as discussed in the main text.

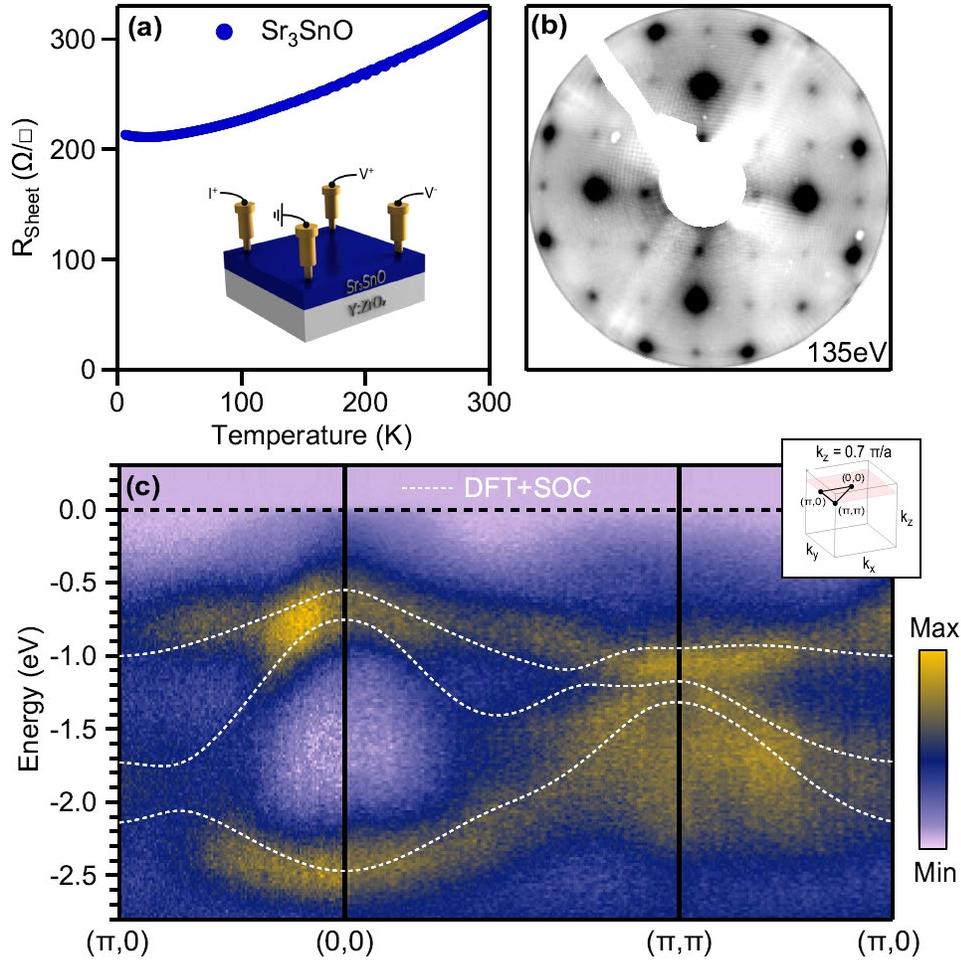

**Fig. 3** Electronic properties of as-grown $Sr_3SnO$ thin films. **a** The metallic behaviour of $Sr_3SnO$ films is revealed by *in-situ* four-point transport experiments. **b** The LEED pattern shows the $Sr_3SnO$ films to have cubic symmetry. **c** The band structure of $Sr_3SnO$ calculated by DFT and measured by ARPES. The agreement between the theoretical and the experimental results confirms the nontrivial topological nature of the $Sr_3SnO$ film.

Acknowledgements

This work was supported by the National Science Foundation under DMREF Grant DMR-1629270 and AFOSR FA9550-15-1-0334. This work was financially supported by the National Science Foundation [Platform for the Accelerated Realization, Analysis, and Discovery of Interface Materials (PARADIM)] under Cooperative Agreement No. DMR-1539918. This work made use of the Cornell Center for Materials Research (CCMR) Shared Facilities, which are supported through the NSF MRSEC program (No. DMR-1719875). Substrate preparation was performed in part at the Cornell NanoScale Facility, a member of the National Nanotechnology Coordinated Infrastructure (NNCI), which is supported by the NSF (Grant No. ECCS-1542081). First-principles calculations of thermodynamic properties (by S.L.S. and Z.K.L.) were carried out partially on the resources of National Energy Research Scientific Computing Center (NERSC) supported by the Department of Energy Office of Science under contract no. DE-AC02-05CH11231, and partially on the resources of Extreme Science and Engineering Discovery Environment (XSEDE) supported by NSF via grant no. ACI-1548562.

Author information

Contributions

C.B.E., D.G.S. and K.M.S. conceived the project and supervised the experiments. Y.M., A.E. and H.P. fabricated thin-film samples and characterized their structures with XRD. B.F. and C.P. carried out and analysed the *in-situ* transport, LEED, XPS and ARPES measurements. B.P. performed the band structure calculations. S.L.S and Z.K.L. calculated the thermodynamic properties by first-principles and the phase diagram. Y.M. drafted the manuscript, which was discussed and refined by all authors.

Competing interests

The authors declare no competing interests.

Corresponding author

Correspondence to C.B.E. at eom@engr.wisc.edu.

# Supplementary Information for

# Realization of Epitaxial Thin Films of the Topological Crystalline Insulator Sr$_3$SnO

Yanjun Ma[1,2], Anthony Edgeton[2], Hanjong Paik[3], Brendan Faeth[4], Chris Parzyck[4], Betül Pamuk[3], Shun-Li Shang[5], Zi-Kui Liu[5], Kyle M. Shen[4,6], Darrell G. Schlom[6,7] and Chang-Beom Eom[2*]

[1]*Department of Physics and Astronomy, West Virginia University, Morgantown, West Virginia 26506, USA*

[2]*Department of Material Science and Engineering, University of Wisconsin-Madison, Madison, Wisconsin 53706, USA*

[3]*Platform for the Accelerated Realization, Analysis, and Discovery of Interface Materials (PARADIM), Cornell University, Ithaca, New York 14853, USA*

[4]*Laboratory of Atomic and Solid State Physics, Cornell University, Ithaca, New York 14853, USA*

[5]*Department of Materials Science and Engineering, The Pennsylvania State University, University Park, Pennsylvania 16802, USA*

[6]*Kavli Institute at Cornell for Nanoscale Science, Ithaca, New York 14853, USA*

[7]*Department of Material Science and Engineering, Cornell University, Ithaca, New York 14853, USA*

---

[*] Correspondence should be sent to eom@engr.wisc.edu.

**Thermodynamic modeling of the Sr-Sn-O system**

The thermodynamic properties of $Sr_3SnO$ were established and used in conjunction with an existing database (the SGTE substance database, SSUB5[1], employed herein) by consideration of the following reaction,

$$3Sr + SnO = Sr_3SnO \qquad (1)$$

The thermodynamic properties of Sr (with Materials Project ID, mp-76)[2], SnO (mp-2097)[2] and $Sr_3SnO$ (mp-7961)[2] are determined by the quasiharmonic approach in terms of DFT-based first-principles and phonon calculations[3]. The obtained reaction Gibbs energy of $Sr_3SnO$ according to Eq. (1) is $-484483 + 18.61\,T$ (Joule per mole formulae at ambient pressure; $T$ is the absolute temperature).

DFT-based first-principles calculations related to thermodynamic properties were performed by the Vienna *Ab initio* Simulation Package (VASP)[4] with the ion-electron interaction described by the projector augmented wave method[5] and the exchange-correlation functional described by the generalized gradient approximation (GGA)[6]. In the VASP calculations, the reciprocal-space energy integration was performed using the Gauss smearing method for structural relaxations and phonon calculations. Final calculations of total energies were performed by the tetrahedron method incorporating a Blöchl correction[7] with a plane-wave cutoff energy of 520 eV. Other details of the first-principles calculations can be found in the Materials Project database[2]. First-principles phonon calculations were performed using the supercell approach as implemented in the YPHON code[8], using the VASP code again as the computational engine to calculate the force constants. The employed supercells (and *k*-point meshes) contain 32 atoms (7×7×7), 72 atoms (3×3×4), and 40 atoms (5×5×5) for Sr, SnO and $Sr_3SnO$, respectively.

Using the SSUB5 database with the thermodynamic data of $Sr_3SnO$ from the present first-principles and phonon calculations, the pressure versus temperature phase stability diagram was calculated using the Thermo-Calc software[9]; see Fig. 1 in the main text.

**Synthesis and characterization of $Sr_3SnO$ thin films on $LaAlO_3$ (001) substrates**

Figure S1a shows the symmetric $\theta$-$2\theta$ scan for $Sr_3SnO$ grown on $LaAlO_3$ (LAO) substrates. The peak at $2\theta \approx 17.297\,°$ is close to the calculated $2\theta \approx 17.25\,°$ for the $Sr_3SnO$ 001 reflection;[10,11] (Cu $K_{\alpha 1}$ radiation was used for the XRD measurements). Higher order $00\ell$ diffraction peaks from $Sr_3SnO$ were identified as well. In addition, no sign of a nonstoichiometric $Sr_{3-x}SnO$ phase was observed in the $\theta$-$2\theta$ scan[11,12]. By fitting the *d*-spacing values calculated for all $00\ell$ peaks with the Nelson-Riley function[13], we find that the extrapolated out-of-plane lattice constant is $c \approx 5.144$ Å. This value is about 0.09% larger than the reported lattice constants of bulk single crystals ($a_{Sr_3SnO} = 5.1394$ Å)[10,11].

Since $a_{Sr_3SnO} = 5.1394$ Å and the pseudocubic lattice parameter of LaAlO$_3$[14], $a_{LAO} = 3.790$ Å, the lattice spacing along the [110] direction of LAO $d_{[110]}^{LAO} \approx 5.360$ Å is close to the lattice parameter of Sr$_3$SnO along the [100] direction; the lattice mismatch is 4.3%. The $\phi$-scan in Fig. S2c confirms the in-plane reorientation, giving an epitaxial relationship of $[110]_{Sr_3SnO} \parallel [100]_{LAO}$ and $(001)_{Sr_3SnO} \parallel (001)_{LAO}$ for the Sr$_3$SnO/LaAlO$_3$ structure.

To investigate the phase purity, we performed asymmetric $\theta$-$2\theta$ scans for the films grown on LAO substrates. As discussed in the main text (see Fig. 2e, f), the single peak in the $\theta$-$2\theta$ scan for the Sr$_3$SnO 202 and 113 reflections, respectively, confirms that our films are free from SrO.

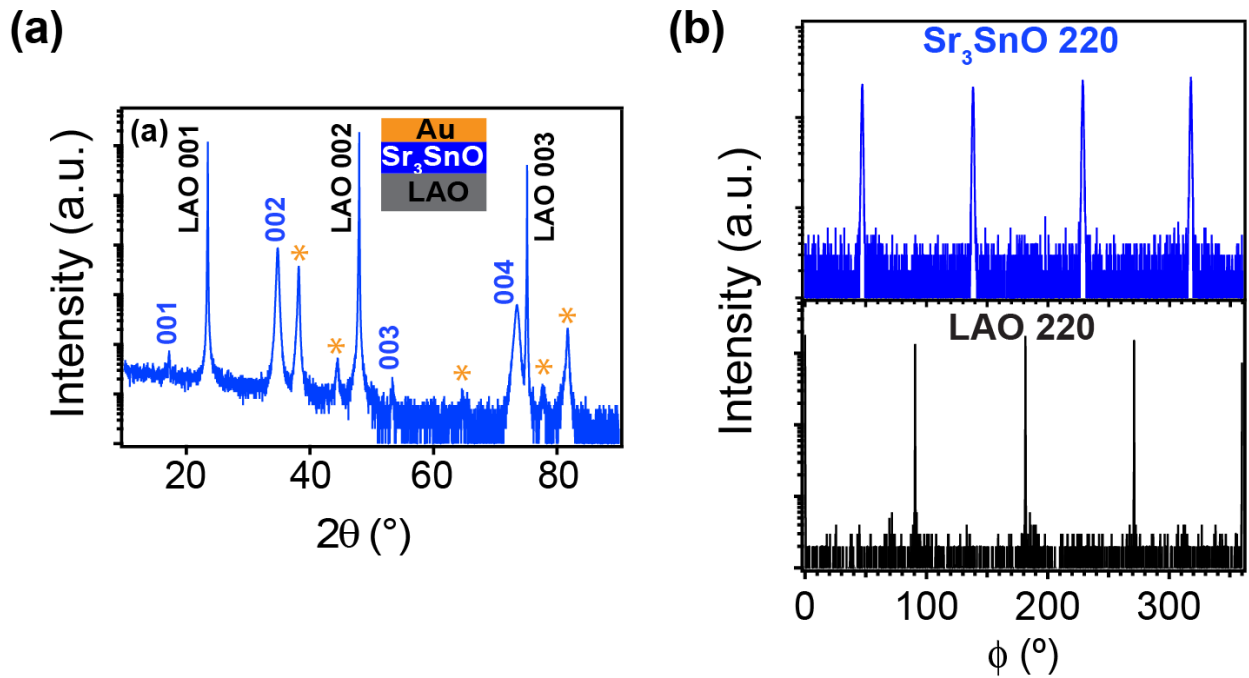

**Figure S1** Characterization of Au-capped Sr$_3$SnO thin films grown on LaAlO$_3$ (001) substrates by *ex-situ* $\theta$-$2\theta$ scan **a** and $\phi$-scan **b**. Asterisks represent the peaks from the Au capping layer.

**Control of the composition and the surface roughness by deposition temperature**

The chemical composition and the surface topography of the as-grown samples sensitively depend on the growth temperature $T$ (Fig. S2). When the synthesis temperature is too low, the SrO phase is formed as demonstrated by *ex-situ* $\theta$-$2\theta$ scans (Fig. S2e). When the substrate temperature is too high, the XRD peaks that can be attributed to SrSn have been identified (Fig. S2f). If the deposition temperature is set in the right range, the desired Sr$_3$SnO structure can form, but with a rough surface, as evidenced by the spotty RHEED patterns (Fig. S2b). We note that similar patterns have been reported previously for BaSnO$_3$ films[15,16]. The single-crystal Sr$_3$SnO thin films with a surface sufficiently smooth for ARPES characterization can be achieved by slowly ramping up the growth temperature until streaky patterns are observed in RHEED (Fig. S2c).

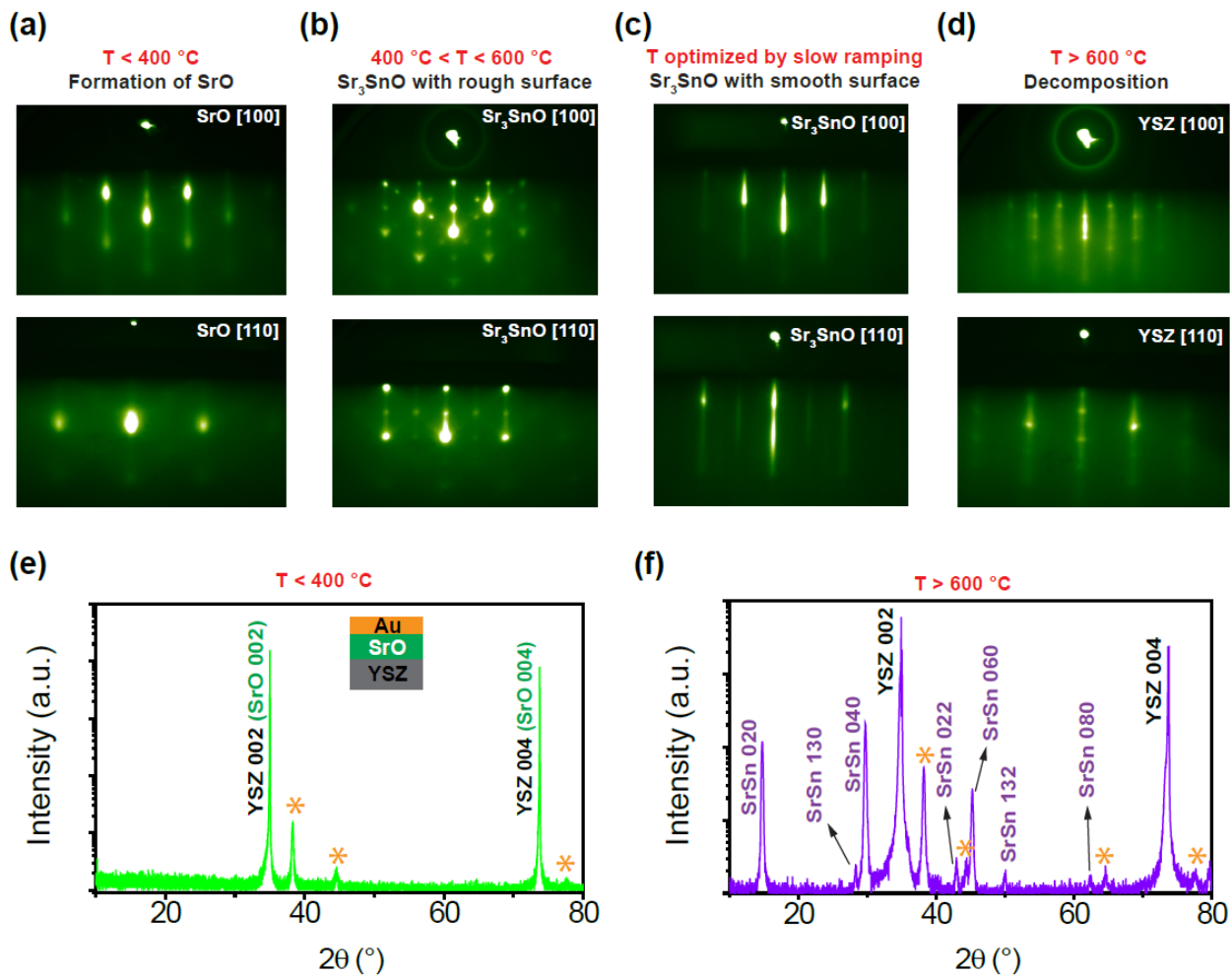

**Figure S2** Structural characterization of samples produced at various temperatures. **a – d** *In-situ* RHEED images along the azimuths indicated for the as-grown films at different deposition temperatures. **e** $\theta$-$2\theta$ scan of a sample generated at low temperature ($T < 500$ °C) shows only the SrO phase. **f** $\theta$-$2\theta$ scan of a sample deposited at high temperatures ($T = 650$ °C) shows reflections that can be assigned to the SrSn compound. Asterisks in both **e** and **f** represent the reflections from the Au capping layers.

**X-ray photoemission spectroscopy measurement**

The chemistry of the synthesized material has been analysed by measuring *in-situ* X-ray photoemission spectroscopy (XPS) on both Sr$_3$SnO samples and a reference SrSnO$_3$ film (Fig. S3). The valence state of Sr in Sr$_3$SnO has been attributed to be +2 because in Fig. S3a, Sr $3p_{1/2}$ and $3p_{3/2}$ peaks from Sr$_3$SnO were found to be close to the ones from SrSnO$_3$ with about $\Delta E_{binding} \approx$ 0.97 eV difference between them. Although the exact chemical shift of $Sn^{4-}$ is currently unknown, the unusual valence state of Sn in $Sr_3SnO$ was manifested by the substantial shift ($\Delta E_{binding} \approx 3.7$ eV) of the Sn $3d$ peaks relative to the reference points. In addition, the broad feature at the higher binding energies have been identified in both the Sn $3d_{3/2}$ and $3d_{5/2}$ peaks, respectively (Fig. S3b). Similar results have been reported for a sister compound $Ca_3SnO$[17]. The O $1s$ state in $SrSnO_3$ showed a single peak, however, the same experiment on $Sr_3SnO$ yielded double peaks. While the peak around the binding energy $E_{binding} \approx 532.5\ eV$ could be ascribed to the oxygen on the surface as discussed before[17], the origin of the peak at $E_{binding} \approx 529$ eV requires further investigation.

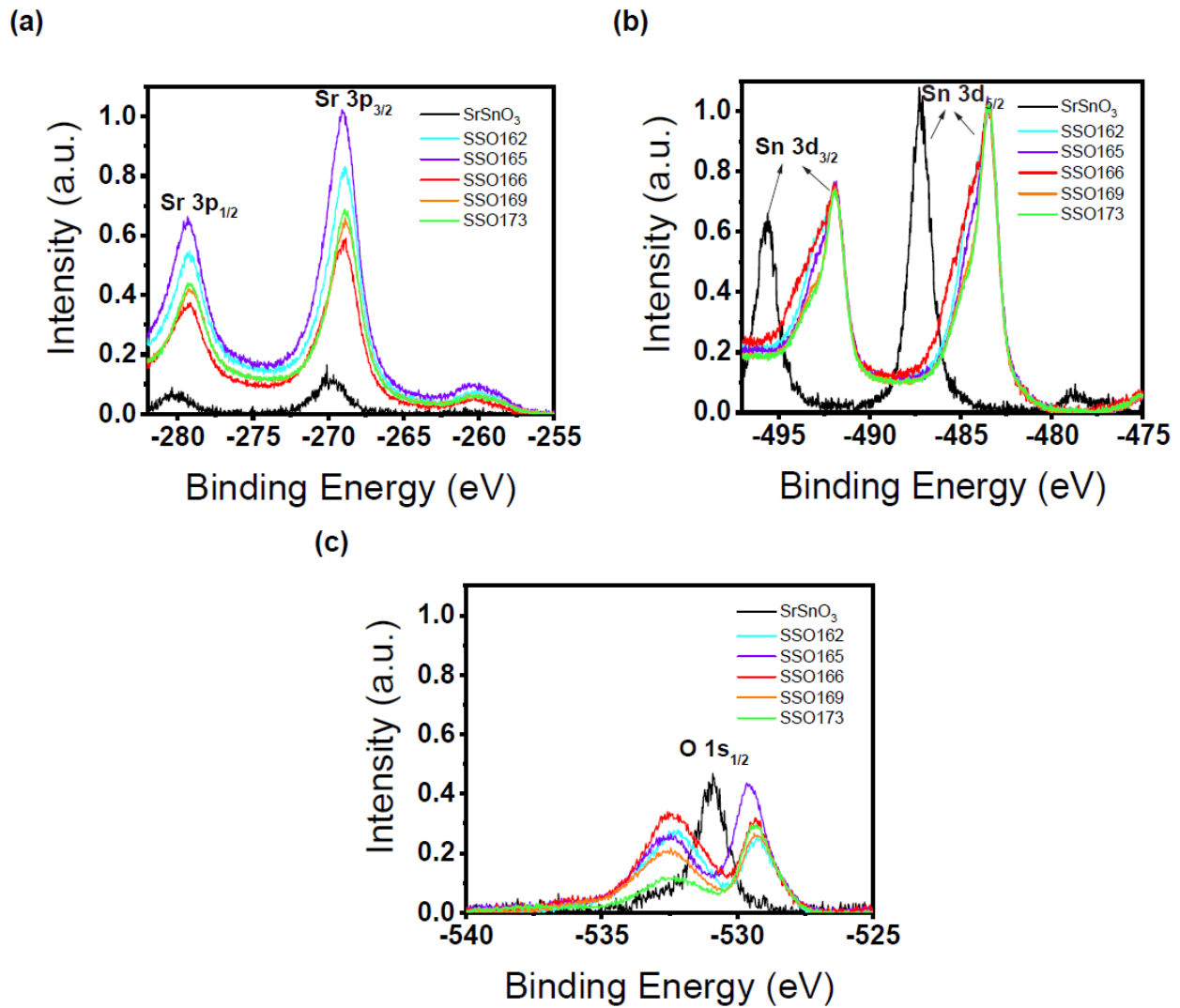

**Figure S3** X-ray photoemission spectroscopy on $Sr_3SnO$ films and a $SrSnO_3$ film reference sample. **a** The Sr $3p$ peaks from $Sr_3SnO$ structure are close to the ones from $SrSnO_3$ with a difference of about 0.95 eV. **b** The Sn $3d$ peaks from $Sr_3SnO$ are shifted relative to the ones from $SrSnO_3$ by approximately 3.7 eV. **c** The O $1s$ state in $Sr_3SnO$ exhibits double-peak feature while the O $1s$ state in $SrSnO_3$ shows a single peak.

**In-situ transport measurements**

As-grown $Sr_3SnO$ thin films characterized by four-point *in-situ* transport measurement showed metallicity between 10 K and 300 K, although the actual sheet resistance varied from sample to sample. Below 10 K, some samples exhibited signatures of localization.

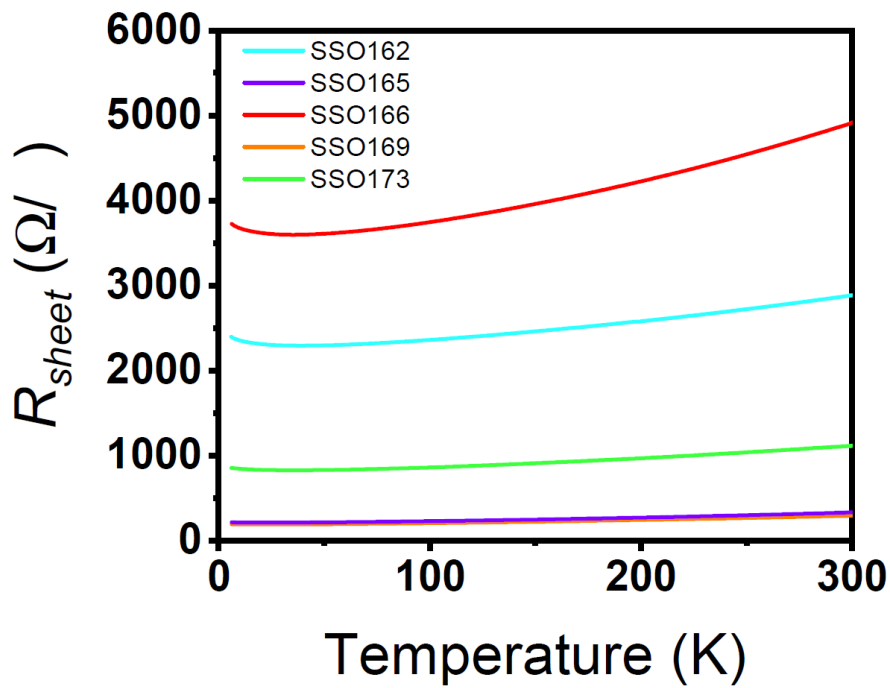

**Figure S4** *In-situ* four-point transport characterization on various $Sr_3SnO$ thin film samples. The metallic nature of $Sr_3SnO$ was revealed for the measured samples.

## Low-Energy Electron Diffraction (LEED) analysis

By carefully analyzing the LEED pattern, we uncover the cubic symmetry for the as-grown $Sr_3SnO$ films with a lattice constant of $a \approx 5.14$ Å (the value in the literature is $a = 5.1394$ Å). This demonstrates that the synthesized samples are low in structural defects, which is corroborated by the fact that the measured LEED result matches the simulated electron diffraction very well.

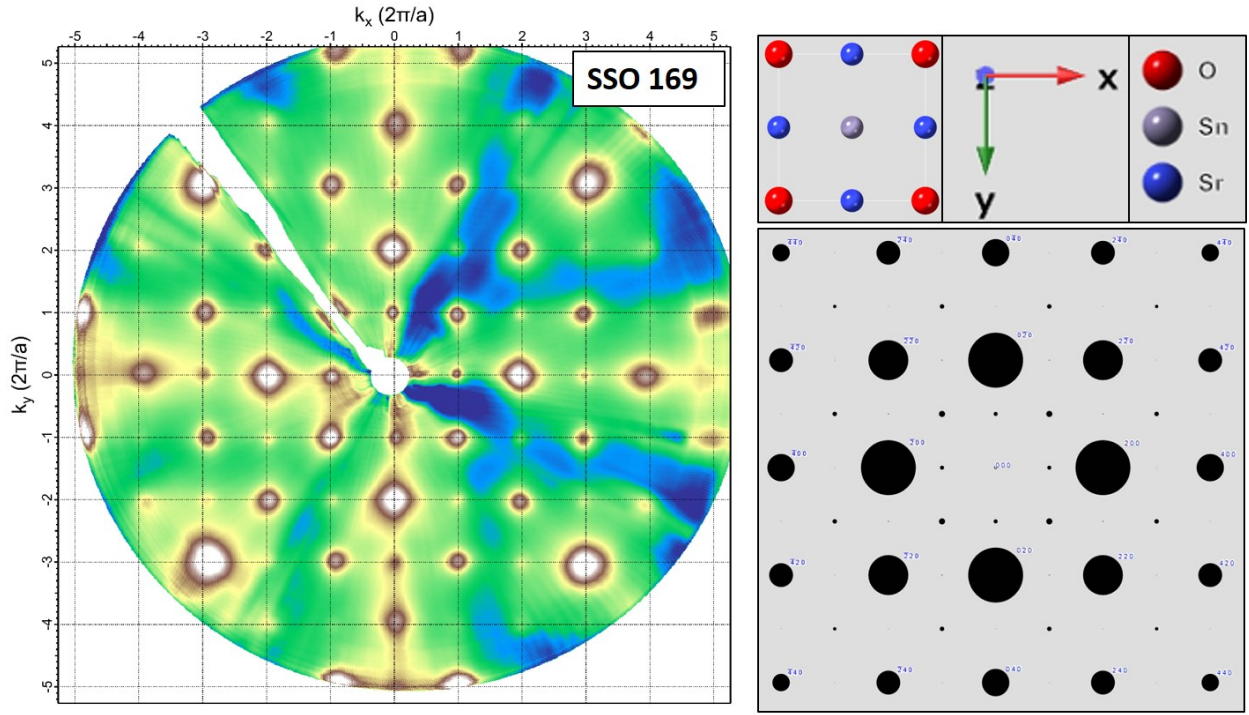

**Figure S5** Low Energy Electron Diffraction analysis. The measured results (left panel) agree very well with the simulated pattern (right panel) confirming that the structural and chemical defects in the as-grown films were minimized.

**Raw Angle-Resolved Photoemission Spectroscopy (ARPES) Spectra**

The following figure shows the raw data of the ARPES experiment on the same sample presented in the Fig. 3 of the main text. Without any post processing, the band structure was already distinguishable from the background noise.

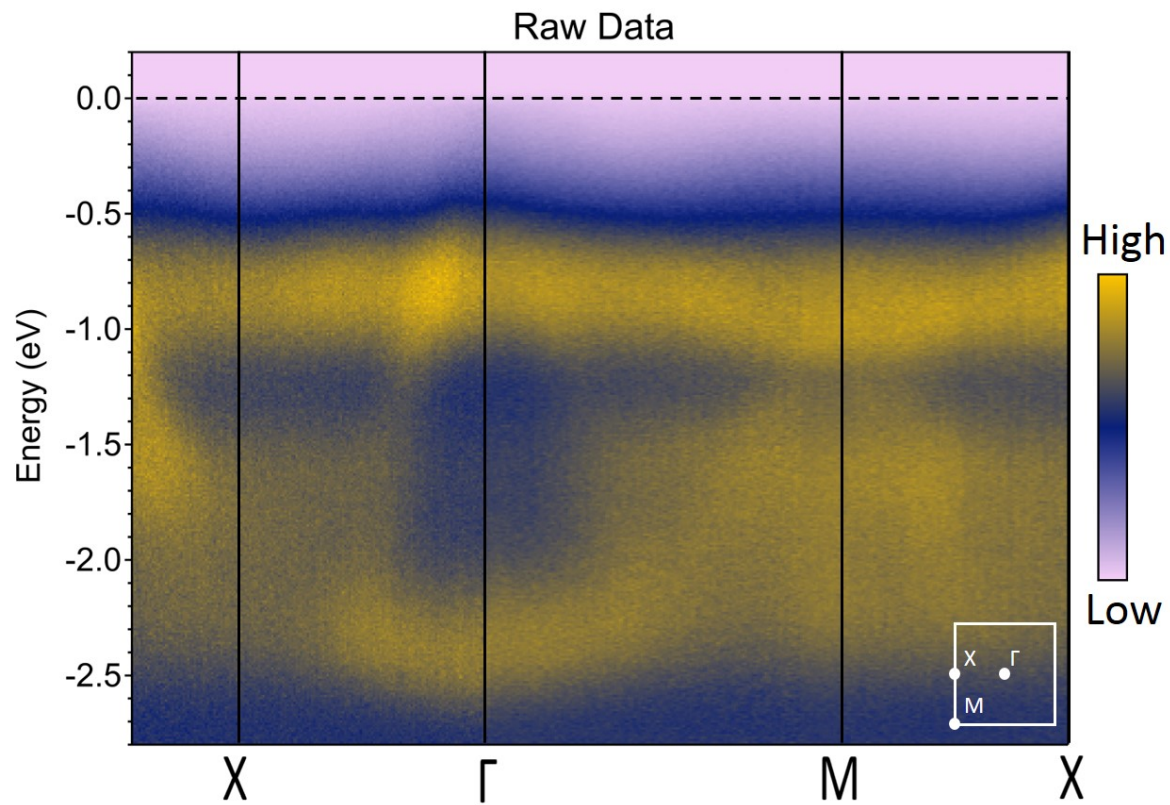

**Figure S6** The raw angle-resolved photoemission spectroscopy (ARPES) result without any post processing. This measurement was done on the same sample as discussed in Fig. 3 of the main text.

## Determination of $k_z$ in Angle-Resolved Photoemission Spectroscopy (ARPES)

The best fitting of our ARPES results with the theoretical calculations was achieved at an out-of-plane momentum of $k_z = 0.7\,\pi/c$ with the computed electronic structure shifted by 0.6 eV (See Fig. S7).

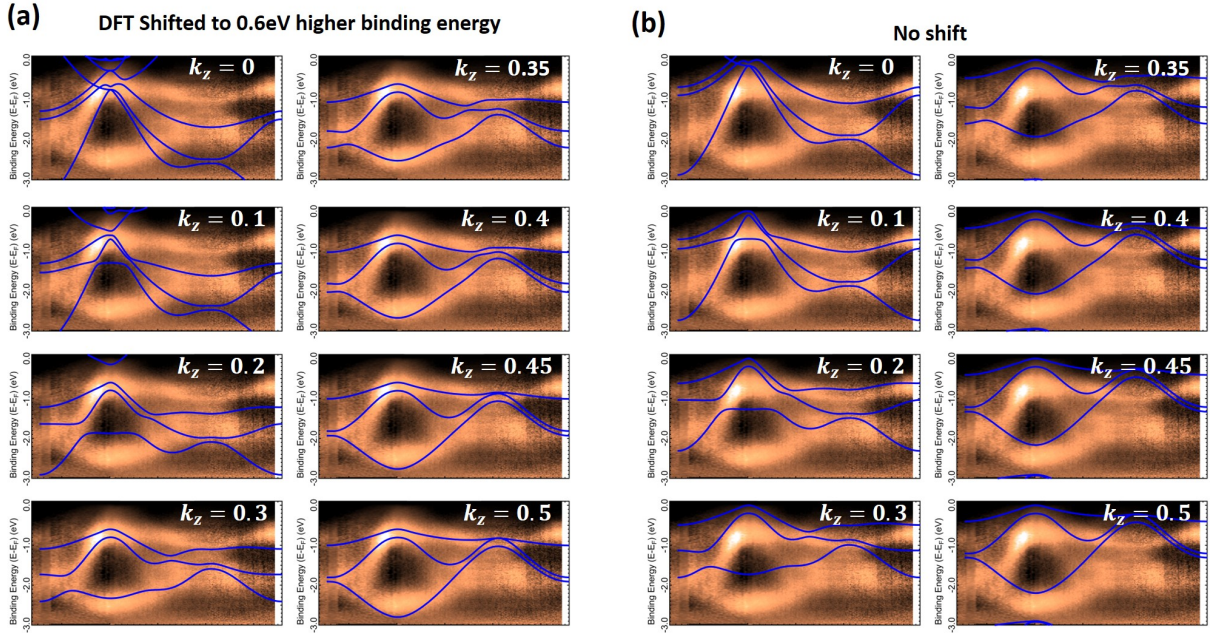

**Figure S7** Fitting of the ARPES data for various $k_z$ values with and without having the calculated band structure shifted (**a** and **b**, respectively).

## DFT-based electronic structure calculations

The electronic band structure and density of states calculations of $Sr_3SnO$ were performed using first-principles density functional theory (DFT) within the GGA as implemented in the PBEsol functional[18] with spin-orbit coupling (SOC). We employed an energy cutoff of 450 eV, electronic momentum $k$-point meshes of $16 \times 16 \times 16$, Methfessel-Paxton smearing of 0.02 eV and energy tolerance of $10^{-8}$ eV for the total energy convergence, at the relaxed lattice parameter of 5.097 Å with force tolerance of $10^{-3}$ Å/eV. The Sumo package[19] was used for plotting the electronic structure.

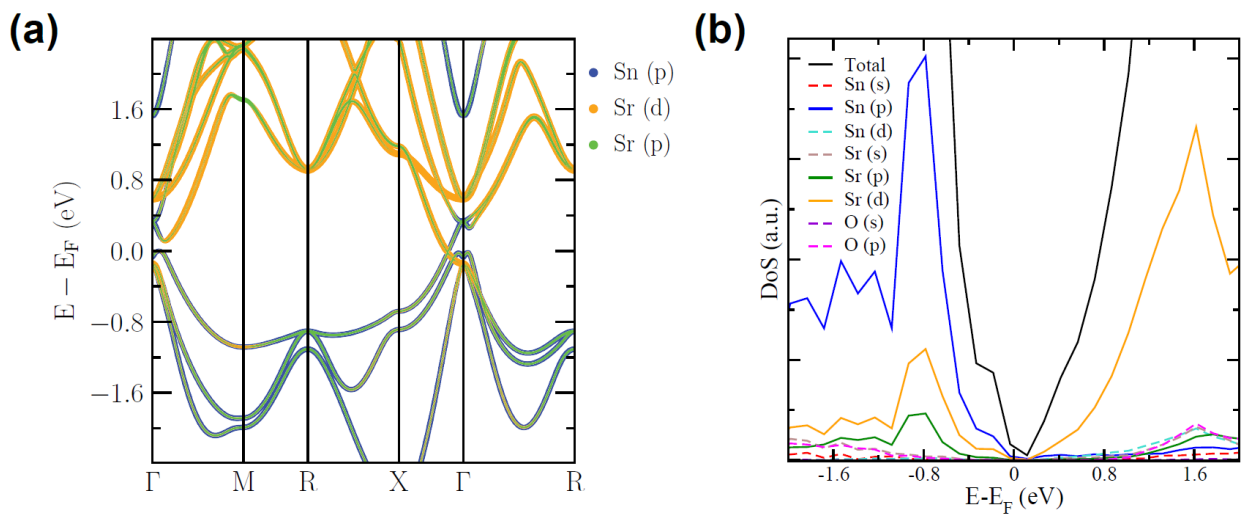

**Figure S8** Band structure and density of states calculations for $Sr_3SnO$. **a** The orbital-projected electronic band structure at the high-symmetry points of the Brillouin zone. The color scale shows the weight of the orbital on the band. **b** The projected density of states (DoS) with Fermi level shifting to 0 eV. It is clear that at the Dirac point, there is a strong hybridization of occupied Sn $5p$ states and unoccupied Sr $4d$ states, consistent with the previous tight-binding calculations[12].